\documentclass[journal]{IEEEtran}
\usepackage{amsmath}
\usepackage{mathrsfs}
\usepackage{pifont}
\usepackage{bbding}
\usepackage{tipa}
\usepackage{amsmath,epsfig}
\usepackage{stmaryrd}
\usepackage{amssymb}
\usepackage{amsfonts}
\usepackage{epic}
\usepackage{graphicx}
\usepackage{curves}
\begin{document}
\newtheorem{theorem}{Theorem}
\newtheorem{proposition}{Proposition}
\newtheorem{definition}{Definition}
\newtheorem{lemma}{Lemma}
\newtheorem{corollary}{Corollary}
\newtheorem{remark}{Remark}
\newtheorem{construction}{Construction}

\newcommand{\supp}{\mathop{\rm supp}}
\newcommand{\sinc}{\mathop{\rm sinc}}
\newcommand{\spann}{\mathop{\rm span}}
\newcommand{\essinf}{\mathop{\rm ess\,inf}}
\newcommand{\esssup}{\mathop{\rm ess\,sup}}
\newcommand{\Lip}{\rm Lip}
\newcommand{\sign}{\mathop{\rm sign}}
\newcommand{\osc}{\mathop{\rm osc}}
\newcommand{\R}{{\mathbb{R}}}
\newcommand{\Z}{{\mathbb{Z}}}
\newcommand{\C}{{\mathbb{C}}}

%=======================================================title==============================================================================
\title{Power Allocation in the High SNR Regime for A Multicast Cell with Regenerative Network
Coding}
%=======================================================author information=================================================================
\author{{Jun~Li and}
        Wen~Chen,~\IEEEmembership{Member,~IEEE}
        % <-this % stops a space
\thanks{Jun~Li and Wen~Chen are with the Department of Electronic Engineering, Shanghai Jiaotong University, Shanghai, 200240
PRC. e-mail: \{jleesr80, wenchen\}@sjtu.edu.cn.}.% <-this % stops a space
\thanks{This work is supported by NSF China \#60672067, by PUJIANG Talents \#07PJ4046, and by Huawei
Fund~\#YJCB2008048WL.}}


%Telephone: (800) 555--1212, Fax: (888) 555--1212}


\maketitle

%=======================================================abstract=================================================================
\begin{abstract}
%\boldmath
This letter focuses on power allocation schemes for a basic
multicast cell with wireless regenerative network coding (RNC). In
RNC, mixed signals received from the two sources are jointly decoded
by the relay where decoded symbols are superposed in either the
complex field (RCNC) or Galois field (RGNC) before being
retransmitted. We deduce the optimal statistical channels state
information (CSI) based power allocation and give a comparison
between the two RNCs. When instantaneous CSI is available at each
transmitter, we propose a suboptimal power allocation for RCNC,
which achieves better performance.
\end{abstract}

% IEEEtran.cls defaults to using math in the Abstract.


% \fi
\IEEEpeerreviewmaketitle

%=======================================================section1 introduction=======================================================
\section{Introduction}\label{sec:1}
Recently, how to leverage network coding~\cite{IEEEconf:1} in
wireless networks to improve system capacity has drawn increasing
interest~\cite{IEEEconf:2}-\cite{IEEEconf:6}. However, these works
focus on the multi-access model or unicast model. Since multicast
topology is popular in practical wireless networks, it is desirable
to investigate network coding in a basic wireless multicast cell.

\begin{figure}[!t]
\begin{picture}(80,24)(-6,0)
\put(10,17){\framebox(8,6){$s_1$}}
%====================================================================
\put(60,17){\framebox(8,6){$d_1$}} \put(18,20){\vector(1,0){42}}
\put(35,20.5){\makebox(8,6)[b]{$\hbar_1$}}
%====================================================================
\put(35,8.5){\framebox(8,6){$r$}} \put(18,19.5){\vector(2,-1){17}}
\put(43,11){\vector(2,1){17}}
\put(32,15){\makebox(0,0)[b]{$g_{\scriptscriptstyle{1}}$}}
\put(44,15){\makebox(3,2)[b]{$h_{\scriptscriptstyle{1}}$}}
%====================================================================
\put(10,0.5){\framebox(8,6){$s_2$}}\put(60,0.5){\framebox(8,6){$d_2$}}
\put(18,2.5){\vector(2,1){17}} \put(43,11.5){\vector(2,-1){17}}
%====================================================================
\put(27,8.5){\makebox(0,0)[b]{$g_{\scriptscriptstyle{2}}$}}
\put(50,8.5){\makebox(3,2)[b]{$h_{\scriptscriptstyle{2}}$}}
%====================================================================
\put(18,2.5){\vector(1,0){42}}
\put(35,3){\makebox(8,6)[b]{$\hbar_2$}}
\end{picture} \caption{$2-1-2$ wireless multicast system.} \label{fig1}
\end{figure}
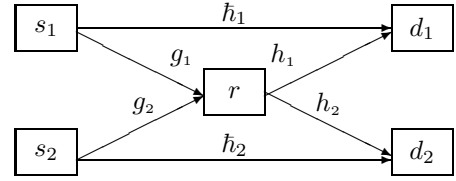

Fig.~\ref{fig1} depicts a basic multicast cell with $2$ sources, $1$
relay and $2$ destinations ($2-1-2$ model). Suppose that both $s_1$
and $s_2$ transmit their messages to the same destination set
$\{d_1, d_2\}$ simultaneously. However, $d_1$ (or $d_2$) is out of
the transmission range of $s_2$ (or $s_1$). The shared relay can
help $s_1$ (or $s_2$) to reach $d_2$ (or $d_1$). When wireless
network coding is applied to the relay, the transmission process
takes two time slots, i.e.,
\\\indent 1. $s_1\rightarrow\{r,~d_1\}$ with $X_{s_1}$; $s_2\rightarrow\{r,~d_2\}$ with $X_{s_2}$,
\\\indent 2. $r\rightarrow\{d_1,~d_2\}$ with $f(X_{s_1}, X_{s_2})$,
\\where $f(\cdot)$ denotes the network coding protocol. In non-regenerative
network coding, the mixed signals from the two sources are not
decoded at the relay before retransmission to the
destinations~\cite{IEEEconf:7}, while in regenerative network coding
(RNC), joint maximum likelihood (ML) decoder is performed at the
relay. Then the decoded symbols are superposed in either the complex
field (RCNC) or Galois field (RGNC) before being retransmitted by
the relay. In this letter, we propose the statistical and
instantaneous CSI based power allocation schemes in the high
signal-to-noise ratio (SNR) regime to improve the system performance
in terms of system frame error probability (SFEP). Throughout this
letter, we use the following notation: $\bar{k}$ denotes the
complementary element of the number $k$ in the set
$\{1\text{,}~2\}$. $\hat{x}$ denotes a decoder's estimate of the
symbol $x$. $\mathcal {E}(\cdot)$ is the statistical expectation.
$z(\rho)\triangleq \mathcal {O}(y(\rho))$, for $y(\rho)>{0}$, means
that there is a positive constants $c$ such that
$|z(\rho)|\leq{c}y(\rho)$ when $\rho$ is large enough.
%=======================================================section2=====================================================================
\section{System Model}\label{sec:2}
Channel coefficients shown in Fig.~\ref{fig1} are assumed to have a
Rayleigh distribution with zero mean and unit variance. The noises
observed by all the receivers are assumed to have a Gaussian
distribution with zero mean and variance $\sigma^2$. We denote $P$
as the average total network transmission power over a time slot.
Then the system SNR is defined as
$\rho\triangleq\frac{P}{\sigma^2}$.

We define $\textbf{x}_{s}\triangleq[x_{s_1},x_{s_2}]$ as a system
frame where $x_{s_k}$ is transmitted by
$s_k\,(k\in\{1\text{,}\,2\})$. The decoded frame to be transmitted
by the relay $r$ in the second time slot is denoted as
$\textbf{x}_{r}=[x_{r_1},x_{r_2}]$. All symbols in $\textbf{x}_{s}$
and $\textbf{x}_{r}$ are $i.i.d$ and selected from the same
$2^R$-QAM constellation set $\mathcal {Q}$ with zero mean and
variance $2P$. The signal received by $d_k$ in the first time slot
is $y_{d_k,1}=\hbar_k\sqrt{\kappa_k}x_{s_k}+v_{d_k,1}$ where
$\kappa_k$ is the power allocation factor (PAF) for $x_{s_k}$.
However, in the second time slot, it is different between the two
protocols, i.e.,
\begin{equation}\label{equ:1}
\begin{split}
&y_{d_k,2}=h_{k}(\sqrt{\tau_1}x_{r_1}+\alpha
\sqrt{\tau_2}x_{r_2})+{v}_{d_k,2}\,\,\,\text{for RCNC},
\\&y_{d_k,2}=h_{k}\sqrt{\tau}x_{r}+v_{d_k,2}\qquad\quad\qquad\qquad\text{for
RGNC},
\end{split}
\end{equation}
where $\alpha=e^{\frac{3j\pi}{4}}$ is the precoder used to achieve
full diversity gain \cite{IEEEconf:3}, $x_r\in\mathcal {Q}$ is the
superposition of $x_{r_1}$ and $x_{r_2}$ in Galois field,
$\sqrt{\tau_k}$ and $\sqrt{\tau}$ are the PAFs of $x_{r_k}$ and
$x_r$ respectively, and $v_{d_k,l}\,(l\in\{1\text{,}\,2\})$ is the
noise observed by $d_k$ in the $l$-th time slot. To compare the two
RNC protocols fairly, we let $\tau=\tau_1+\tau_2$. Then we have
$\mathcal {E}(\tau_1|x_{r_1}|^2+\tau_2|x_{r_2}|^2)=\mathcal
{E}(\tau|x_r|^2)$. Define $\kappa\triangleq\kappa_1+\kappa_2$, and
then $\kappa+\tau=1$. So the total power consumed during a frame
period, i.e., two time slots, is
\begin{equation}\label{equ:2}
\mathcal{E}_{\textbf{x}_s,\textbf{x}_r}\left(\kappa_1|x_{s_1}|^2+\kappa_2|x_{s_2}|^2+\tau|x_{r}|^2\right)=2P.
\end{equation}
Note that ML decoding is performed at all receivers. In RCNC
protocol, if $r$ can successfully decode $\textbf{x}_s$, i.e.,
$\textbf{x}_r=\textbf{x}_s$, then after the second time slot, the
joint ML decoder at $d_k$ is
\begin{equation}\label{equ:3}
\begin{split}
(\hat{x}_{s_1},\hat{x}_{s_2})_{d_k}=&\arg\underset{x_{s_1},x_{s_2}\in\mathcal
{Q}}{\min}\big\{|y_{d_k,1}-\hbar_k\sqrt{\kappa_k}x_{s_k}|^2\\&+|y_{d_k,2}-h_{k}(\sqrt{\tau_1}x_{s_1}+\alpha
\sqrt{\tau_2}x_{s_2})|^2\big\}.
\end{split}
\end{equation}
While in RGNC protocol, $d_k$ decodes $x_{s_k}$ after the first time
slot and decodes $x_r$ after the second time slot since the two
symbols are mutually independent. Then $x_{s_{\bar{k}}}$ can be
worked out by Galois field operation between $x_{s_k}$ and $x_r$.
%=======================================================section3=====================================================================
\section{Power Allocation Schemes}\label{sec:3}
We suppose that when $\textbf{x}_s$ is wrongly decoded by either of
the destinations, a system frame error event (SFEE) occurs. Then
SFEP is defined as the probability of SFEE, which can be expressed
as
$P_{sys}=P_{d_1}(1-P_{d_2})+P_{d_{2}}(1-P_{d_1})+P_{d_{1}}P_{d_{2}}$,
where $P_{d_k}$ is the FEP of $d_k$, i.e., the probability of the
event that $d_k$ wrongly decodes $\textbf{x}_s$. According to the
system model, if $r$ wrongly decodes $\textbf{x}_s$, there is at
least one destination $d_k$ which can not extract the right symbol
$x_{s_{\bar{k}}}$ from $r$ and thus SFEE occurs with probability
$1$. Then $P_{sys}$ is rewritten as
\begin{equation}\label{equ:4}
P_{sys}=P_{r}+(1-P_{r})(P_{d_1|r}+P_{d_2|r}-P_{d_1|r}P_{d_2|r}),
\end{equation}
where $P_{r}$ is the FEP of $r$ and $P_{d_k|r}$ is the FEP of $d_k$
on the condition that $r$ can successfully decode $\textbf{x}_s$. In
the sequel, we will optimize the PAFs of the two RNCs to minimize
$P_{sys}$ according to the CSI available at the transmitters.
%=====================================================subsection A=============================================
\subsection{Statistical CSI based Power Allocation Scheme}
Due to the statistical symmetry of the channel model, the PAFs are
chosen as $\kappa_1=\kappa_2=\frac1{2}\kappa$ and
$\tau_1=\tau_2=\frac1{2}\tau$. To find the optimal relation between
$\kappa$ and $\tau$, we firstly focus on $P_r$. We denote $P_{PE,r}$
as the average pairwise error probability (APEP) of $r$. Since there
are in total $2^{2R}$ codewords, we have $P_{r}=2^{2R}P_{PE,r}$. By
taking expectation with respect to $[g_1\text{,}\,g_2]$, statistical
CSI based $P_{PE,r}$ can be deduced as~\cite{IEEEconf:8}, i.e.,
\begin{equation}\label{equ:5}
P_{PE,r}=\mathcal {E}_{u_{s_1},u_{s_2}}\left\{\frac{\rho^{-1}}{\pi}\int_0^{\frac{\pi}{2}}\left(\frac1{\rho}+\frac{|u_{s_1}|^2+|u_{s_2}|^2}{8\sin^2\theta}\right)^{-1}\,\mathrm{d}\theta\right\},
\end{equation}
where $u_{s_k}=\sqrt{{\kappa}_k/{P}}(x_{s_k}-\hat{x}_{s_k})$ is the
normalized decoding error of the symbol $x_{s_k}$. When $\rho$ is
large, we omit the factor $\frac1{\rho}$  inside the integral in
(\ref{equ:5}). Then $P_r$ can be approximated as
\begin{equation}\label{equ:6}
P_{r}\approx2^{2R}\mathcal
{E}_{u_{s_1},u_{s_2}}\left\{\frac{2\rho^{-1}}{|u_{s_1}|^2+|u_{s_2}|^2}\right\}.
\end{equation}
Next, we focus on $P_{d_k|r}$ of the two protocols. In RCNC, joint
ML decoding is performed at $d_k$ shown as (\ref{equ:3}). Since
$x_{s_k}$ can achieve more diversity gain than $x_{s_{\bar{k}}}$,
then in the high SNR regime, $P_{d_k|r}$ is dominated by the
probability of the event that $x_{s_k}$ is successfully decoded but
$x_{s_{\bar{k}}}$ is wrongly decoded. So
\begin{equation}\label{equ:7}
P_{d_k|r}^{RCNC}\approx2^R\mathcal
{E}_{u_{r_{\bar{k}}}}\left\{\frac{2\rho^{-1}}{|u_{r_{\bar{k}}}|^2}+\mathcal
{O}(\rho^{-2})\right\},
\end{equation}
where
$u_{r_{\bar{k}}}=\sqrt{{\tau}_{\bar{k}}/{P}}(x_{s_{\bar{k}}}-\hat{x}_{s_{\bar{k}}})$
is the normalized decoding error of the symbol $x_{s_{\bar{k}}}$.
While in RGNC, $x_{s_k}$ and $x_{r}$ are mutually independent and
received by $d_k$ in time division channels. So when $\rho$ is large
enough, we get
\begin{equation}\label{equ:8}
P_{d_k|r}^{RGNC}\approx2^R\mathcal
{E}_{u_{s_k},u_r}\left\{\frac{2\rho^{-1}}{|u_{s_{k}}|^2}+\frac{2\rho^{-1}}{|u_{r}|^2}\right\},
\end{equation}
where $u_{r}=\sqrt{{\tau}/{P}}(x_{r}-\hat{x}_{r})$ is the normalized
decoding error of the symbol $x_{r}$. In the sequel, we give the
statistical CSI based power allocation of the two protocols
respectively.
\begin{theorem}\label{theo1}
When $\rho$ is large enough, the optimal statistical CSI based
optimal power allocation is to choose the PAF $\kappa$ as
\begin{equation}\label{equ:9}
\kappa^c=\frac{\sqrt{2^{R-2}}}{\sqrt{2^{R-2}}+1}\text{for RCNC,~}
\kappa^g=\frac{\sqrt{2^{R-1}+2}}{\sqrt{2^{R-1}+2}+1}\text{for RGNC}
\end{equation}
\end{theorem}

\emph{Proof:} When $\rho$ is large enough, we rewrite (\ref{equ:4})
as $P_{sys}\approx{P}_{r}+P_{d_1|r}+P_{d_2|r}$. Since $\mathcal
{E}(|x_{s_k}-\hat{x}_{s_k}|^2)=\mathcal
{E}(|x_{r}-\hat{x}_{r}|^2)=4P$, the expectations of the decoding
error $\mathcal {E}(|u_{s_k}|^2)=2\kappa$, $\mathcal
{E}(|u_{r_k}|^2)=2\tau$ and $\mathcal {E}(|u_{r}|^2)=4\tau$. Then we
approximate the $P_{sys}$ of the two protocols by their upper
bounds, i.e.,
\begin{equation}\label{equ:10}
\begin{split}
P_{sys}^{RCNC}&\approx2^{2R}\frac{2\rho
^{-1}}{4\kappa}+2\cdot2^R\frac{2\rho^{-1}}{2\tau}\\&=2^{R}\rho^{-1}\left(\frac{2^{R-1}}{\kappa}+\frac{2}{\tau}\right).
\end{split}
\end{equation}
So the optimal power allocation of RCNC can be worked out by
minimizing $\left(\frac{2^R}{2\kappa}+\frac{2}{\tau}\right)$ subject
to the power constraint $\kappa+\tau=1$. On the other hand, in RGNC
protocol, we have
\begin{equation}\label{equ:11}
\begin{split}
P_{sys}^{RGNC}&\approx2^{2R}\frac{2\rho
^{-1}}{4\kappa}+2\cdot\left(2^R\frac{2\rho^{-1}}{2\kappa}+2^R\frac{2\rho^{-1}}{4\tau}\right)\\&=2^{R}\rho^{-1}\left(\frac{2^{R-1}+2}{\kappa}+\frac1{\tau}\right).
\end{split}
\end{equation}
By minimizing $\left(\frac{2^{R-1}+2}{\kappa}+\frac1{\tau}\right)$
subject to the power constraint $\kappa+\tau=1$, we get the optimal
power allocation of RGNC.~~~{\ding{110}}
%=====================================================subsection B=============================================
\subsection{Instantaneous CSI based Power Allocation}
If instantaneous CSI is available at all transmitters, PAFs can be
further optimized. In the first time slot, we focus on guaranteeing
the quality of both $s\rightarrow{r}$ channels to minimize $P_r$,
which is a multi-access channel model. According
to~\cite{IEEEconf:9}, we suppose that each source splits its power
into $M$ pieces, i.e., $2\kappa_{k}\rho=M\triangle{\rho}_k$. Two
sources alternatively pour one piece of their power into the
channels and gain the rate growth $\triangle{R}(s_{k}^m)$ in the
$m$-th round. Let $\triangle{\rho}_{k}\rightarrow{0}$. Then we have
$\triangle{R}(s_{k}^m)=\frac1{2}|g_{k}|^2\triangle{\rho_k}\eta_{m}$,
where
$\eta_{m}={1}/({1+m{\sum}_{j=1}^2{|g_{j}|^2\triangle\rho_{j}}})$.
When joint ML decoding is performed at $r$,
$\triangle\rho_{\bar{k}}$ can be replaced by
$\frac{\kappa_{\bar{k}}}{\kappa_{k}}\triangle\rho_{k}$, i.e.,
\begin{equation}\label{equ:12}
\begin{split}
I(s_k;r|g_1,g_2)&=\int_0^{2\kappa_{k}\rho}\frac{\frac1{2}|g_{k}|^2\,\mathrm{d}\rho_k}{1+(|g_{k}|^2+\frac{\kappa_{\bar{k}}}{\kappa_{k}}|g_{\bar{k}}|^2)\rho_{k}}
\\&=\frac{\kappa_{k}|g_{k}|^2}{\kappa_{k}|g_{k}|^2+\kappa_{\bar{k}}|g_{{\bar{k}}}|^2}I(s_k,s_{\bar{k}};r|g_1,g_2).
\end{split}
\end{equation}
Let $I(s_k;r|g_1,g_2)=I(s_{\bar{k}};r|g_1,g_2)$ to guarantee the
quality of the worse channel. Then we the power allocation between
the two sources as
$\kappa_k={\kappa|g_{\bar{k}}|^2}/({|g_k|^2+|g_{\bar{k}}|^2})$.
Moreover, the phase of each $s\rightarrow{r}$ channel is
pre-equalized to ensure the coherent superposition of the two
signals. Then we focus on the instantaneous CSI based power
allocation in RCNC. Due to space limitations, the discussion on RGNC
is omitted.
\begin{theorem}\label{theo2} A suboptimal instantaneous CSI based power allocation for RCNC is to
choose the PAFs as
\begin{equation}\label{equ:13}
\kappa^c=\frac{\sqrt{\eta2^{R-1}}}{\sqrt{\eta2^{R-1}}+1},\,\tau^c=\frac1{\sqrt{\eta2^{R-1}}+1},\,\tau_k^c=\frac{\tau{|h_k|}}{|h_k|+|h_{\bar{k}}|},
\end{equation}
where
$\eta=\frac{|h_1h_2|^2(|g_1|^2+|g_2|^2)}{|g_1g_2|^2(|h_1|+|h_2|)^2}$.
\end{theorem}

\emph{Proof:} Since the instantaneous CSI based SFEP can not be
exactly worked out, we give a suboptimal method by replacing the
statistical SNR in (\ref{equ:10}) with the instantaneous SNR. Then
the suboptimal power allocation is to minimize
$\left(\frac{2^{R-1}}{\kappa{|g|^2}}+\frac1{\tau_1{|h_2|^2}}+\frac1{\tau_2{|h_1|^2}}\right)$
subject to the power constraint $\kappa+\tau_1+\tau_2=1$, where
$|g|^2={|g_1g_2|^2}/({|g_1|^2+|g_2|^2})$. Then we complete the
proof.~~~~~~~~~~~~~~~~~~~~~~~~~~~~~~~~~~~~~~~~~~~~~~~~~~~{\ding{110}}

%The right hand side of (\ref{equ:12}) is deduced under the
%assumption that symbols in $s\rightarrow{d}$ link are not multiplied
%by $\textbf{\textrtailn}_{s_k}$, while in the real multicast system
%with DP, both $s\rightarrow{d}$ and $s\rightarrow{r}$ link are
%effected by the $\textbf{\textrtailn}_{s_k}$ which can bring more
%diversity gain in $s\rightarrow{d}$ link, thus get lower SFEP.

%=======================================================section4=====================================================================
\section{Analysis and Numerical Results}\label{sec:4}
Consider the conventional scheme without network coding where all
signals are transmitted in time division (TD) channels. The
transmission of $\textbf{x}_s$ should take $4$ time slots and thus
consumes more power and time slots than that of network coding
schemes. So network coding schemes outperform the conventional
scheme in terms of system throughput. This issue has been thoroughly
investigated in previous works~\cite{IEEEconf:2}-\cite{IEEEconf:6}.

\begin{figure}[!t]
\centering
\includegraphics[width=3.5in,angle=0]{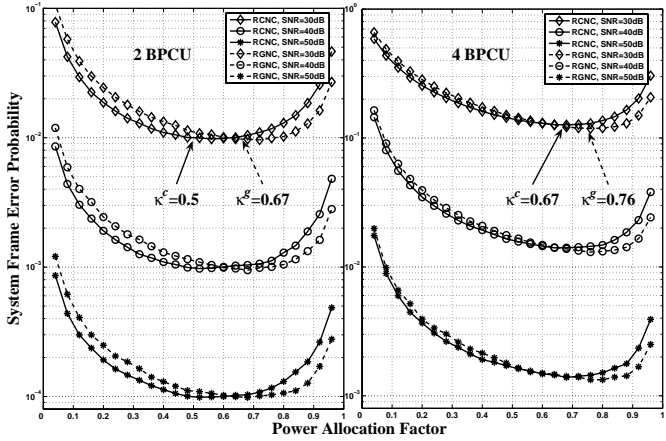}
\caption{SFEP with different statistical CSI based PAS for RCNC and
RGNC protocols respectively. The horizontal axis represents the PAF
$\kappa$.
%We consider $2$ BPCU and $4$ BPCU with
%QAM modulation. Abscissa denotes the value of power allocation
%factor $\kappa$.
} \label{fig2}
\end{figure}

In our Monte-Carlo simulations, decoding algorithm and system model
are both selected as that in section~\ref{sec:2}. Each SFEP value is
simulated by $10^6$ $i.i.d$ frames. Fig.~\ref{fig2} shows the SFEP
curves with different values of PAF $\kappa$ where statistical CSI
(SCSI) based power allocation of the two protocols are considered.
Since the optimal power allocation given by \emph{Theorem
\ref{theo1}} are related to $R$, we consider two QAM modulation
schemes, i.e., $2$~bit per-channel use (BPCU) and $4$ BPCU.
According to (\ref{equ:9}), in $2$~BPCU scenario, the optimal power
allocation for RCNC is to choose $\kappa^c=1/2$ and for RGNC is
$\kappa^g=2/3$, while in $4$~BPCU scenario, the optimal power
allocation is to choose $\kappa^c=2/3$ and
$\kappa^g=\sqrt{10}/(\sqrt{10}+1)\approx0.76$ for the two protocols
respectively. Fig.~\ref{fig2} shows that \emph{Theorem \ref{theo1}}
accurately predicts the SCSI based optimal power allocation.
\begin{figure}[!t]
\centering
\includegraphics[width=3.5in,angle=0]{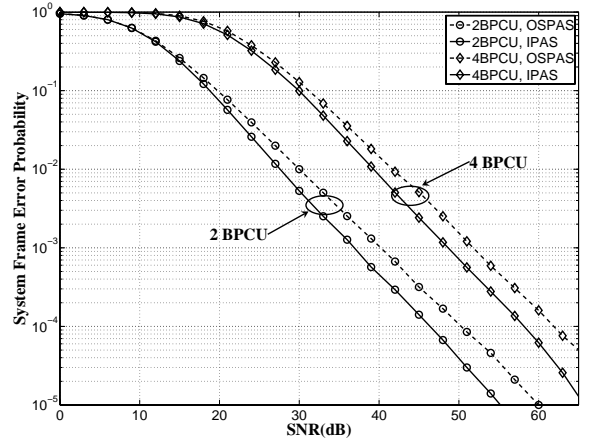}
\caption{Optimal statistical CSI based PAS (OSPAS) vs instantaneous
CSI based PAS (IPAS) with $2$ BPCU and $4$ BPCU respectively in
RCNC.} \label{fig3}
\end{figure}

\emph{Theorem \ref{theo1}} also provides a comparison between the
two protocols. Note that in multi-access model~\cite{IEEEconf:5} and
unicast model, Galois field network coding outperforms the complex
field network coding. However, this is not always true in our
multicast system. We compare the performance of the two protocols
according to (\ref{equ:10}) and (\ref{equ:11}). Let
$P_{sys}^{RCNC}=P_{sys}^{RGNC}$, which means that the two protocols
have the same system performance. Then we get $\kappa={2}/{3}$ and
$\tau=1/{3}$. If PAF is chosen as $\kappa<2/3$ (or $\kappa>2/3$), we
have $P_{sys}^{RCNC}<P_{sys}^{RGNC}$ (or
$P_{sys}^{RCNC}>P_{sys}^{RGNC}$). Then RCNC is better (or worse)
than RGNC with the performance difference
$\Delta{P}_{sys}=\left|2^R\rho^{-1}(\frac{2}{\kappa}-\frac1{\tau})\right|$.
Fig.~\ref{fig2} proves our predictions.

Fig.~\ref{fig3} compares the optimal statistical CSI based power
allocation scheme (OSPAS) with the instantaneous CSI based power
allocation scheme (IPAS) in RCNC. $2$ BPCU and $4$ BPCU are
respectively considered. With the instantaneous CSI at each
transmitter, IPAS drastically outperforms the OSPAS.

\section{Conclusion}
In this letter, we analyze the power allocation schemes for RCNC and
RGNC protocols in $2-1-2$ multicast system. In the high SNR regime,
the optimal statistical CSI based power allocation is proposed by
\emph{Theorem \ref{theo1}} in terms of SFEP. According to
\emph{Theorem} \ref{theo1}, we also give a comparison of the two
RNCs. When instantaneous CSI is available at transmitters, the
suboptimal but simple power allocation proposed by \emph{Theorem
\ref{theo2}} can further improve the system performance of RCNC.



%\bibliographystyle{IEEEtran}


\end{document}